\documentclass[aps,prb,showpacs,twocolumn,showpacs]{revtex4-1}

\usepackage{graphicx}
\usepackage{amsmath}
\usepackage{amssymb}
\usepackage{wasysym}
\usepackage{color}

\definecolor{lila}{rgb}{0.5,0,1}
\newcommand{\bnen}{\begin{equation}}
\newcommand{\eden}{\end{equation}}
\newcommand{\bean}{\begin{eqnarray}}
\newcommand{\eean}{\end{eqnarray}}

\newcommand{\bna}{\begin{array}}
\newcommand{\eda}{\end{array}}

\begin{document}

\title{Momentum distribution functions in a one-dimensional extended periodic
Anderson model}

\author{I. Hagym\'asi}
\author{J. S\'olyom}
\author{\"O. Legeza}

\affiliation{Strongly Correlated Systems "Lend\"ulet" Research Group, Institute for Solid State
Physics and Optics, MTA Wigner Research Centre for Physics, Budapest H-1525 P.O. Box 49, Hungary
}

\date{\today}

\begin{abstract}
We study the momentum distribution of the electrons in an extended periodic
Anderson model, where the interaction, $U_{cf}$, between itinerant and localized electrons is taken
into
account. In the symmetric half-filled model, due to the increase of the interorbital interaction,
the $f$ electrons become more and more delocalized, while the itinerancy of conduction
electrons decreases. Above a certain value of $U_{cf}$ the $f$ electrons become again localized
together with the conduction electrons. In the less than half-filled case, we observe that $U_{cf}$
causes strong correlations between the $f$ electrons in the mixed valence regime.
\end{abstract}

\pacs{71.10.Fd, 71.27.+a, 75.30.Mb}

\maketitle

\section{Introduction}
Heavy-fermion and mixed valence systems are still active research fields in spite of the major
achievements of the past few decades.\cite{Hewson:book,Patrik:book} 
The discovery of a new critical
point in the pressure-temperature phase diagram of CeCu$_2$Ge$_2$ and CeCu$_2$Si$_2$ has attracted
much attention both
experimentally\cite{Yuan:science,CeCu2Si2:meres_2,CeCu2SiGe:meres,CeCu2Si2:meres_1,
CeCu2Si2:meres_SC,RueffPRL_meres:cikk} and
theoretically.\cite{Miyake:VMC,Miyake:CVF_1,DMRG:Miyake1,DMRG:Miyake2,Miyake_SC,Miyake:review} It is
believed that the appearance of the new critical point is due to the critical valence fluctuations
of the Ce ion.
The simplest model, which contains the essential physics of rare-earth compounds is the periodic
Anderson model:
\begin{equation}   \begin{split}
     \mathcal{H}_{\rm PAM} = & -t\sum_{<ij>,\sigma}
       \hat{c}_{i\sigma}^{\dagger}
	   \hat{c}^{\phantom \dagger}_{j\sigma}\\
&-V\sum_{j,\sigma}(\hat{f}_{j\sigma}^{\dagger}
	  \hat{c}^{\phantom \dagger}_{j\sigma}
    +\hat{c}_{j\sigma}^{\dagger} \hat{f}^{\phantom \dagger}_{j\sigma}) 
+\varepsilon_f\sum_{j,\sigma}\hat{n}^f_{j\sigma}\\
&+U_f\sum_{j}\hat{n}^f_{j\uparrow}
\hat{n}^f_{j\downarrow}
\label{PAM:Hamiltonian}
	     , 
\end{split}
\end{equation}
where the notation is standard and $W=4t$ is taken as the energy unit.
It is known, however, that 
the mixed-valence regime appears always in this model as a smooth crossover, and valence
fluctuations do not become
critical for any choice of the parameters. A local Coulomb interaction between the conduction and
localized electrons is
needed for the appearance of a sharp transition and critical valence
fluctuations.\cite{Miyake:review} Therefore we consider the following Hamiltonian:
\begin{equation}
 \mathcal{H} =\mathcal{H}_{\rm PAM}+U_{cf}\sum_{j,\sigma,\sigma'}
\hat{n}^f_{j\sigma}\hat{n}^c_{j\sigma'}.
\label{eq:EPAM}
\end{equation}
Previous studies revealed how $U_{cf}$ affects the mixed valence regime and it has been shown that a
first-order valence transition and a quantum critical point may appear due to
$U_{cf}$.\cite{DMRG:Miyake1,Hirashima:cikk,Kubo:GW,Hagymasi:GW} The effect of $U_{cf}$ in the Kondo
regime has been also addressed both in infinite\cite{Kawakami:DMFT_1,Kawakami:DMFT_2} and one
spatial dimensions.\cite{Hagymasi:Ucf} Namely, in infinite dimensions the symmetric model for small
hybridization ($V\ll W$) displays antiferromagnetic
order for small $U_{cf}$ which, however, disappears for large $U_{cf}$ and charge order develops.
In contrast, there is no such phase transition in one dimension due to the enhanced quantum
fluctuations, however, for small and large $U_{cf}$ the spin-spin and density-density correlation
function, respectively exhibits the slowest decay. Between these two regimes there is a narrow
region, where the local singlet formation is significantly enhanced.
\par Our goal in this paper is to investigate the momentum distribution of the electrons in one
dimension. It is known, that in higher dimensions they exhibit a jump at the Fermi momentum, whose
size can be used to extract the energy dependence of the self-energies, from which the many-body
enhancement factor of the effective mass can be obtained. Although in one dimension there is no
such  jump at the Fermi momentum, just a sharp change, they provide direct
information about the spatial distribution of the electrons and the content of conduction and
$f$-electron states in the quasiparticle bands, while the previous quantum information
analysis\cite{Hagymasi:Ucf} gave only an indirect description of these quantities. We
address the question how they are modified by
switching on $U_{cf}$ both in the integer and mixed valence regimes. The density-matrix
renormalization-group algorithm
(DMRG)\cite{White:DMRG1,White:DMRG2,schollwock2005,manmana2005,hallberg2006} is applied, which
allows the accurate determination of ground state properties. We have used the dynamic block-state
selection algorithm\cite{DBSS:cikk1,DBSS:cikk2} in which the threshold value of the quantum
information loss,
$\chi$, is set a priori. We have taken  
$\chi=10^{-5}$. A maximum of 2000 block states is needed to achieve this 
accuracy, and the largest truncation error was in the order of $10^{-6}$. We investigated chains up
to a maximum length $L=80$ with open boundary conditions and performed
8-12 sweeps.
\section{Results at half filling}
The non-degenerate version of the periodic Anderson model
can hold up to $n_{\rm max}=4$ electrons per lattice site, the 
average number of $c$ and $f$ electrons per site, $n^c$ and $n^f$, respectively, 
can vary between zero and two. The filling will refer to the ratio of the total 
electron density per site ($n^c + n^f$).
In what follows we consider the symmetric half-filled model, where $n^f=1$, and calculate the
momentum
distribution of conduction and $f$ electrons which are defined as
\begin{align}
 n^c(k)=\frac{1}{2}\sum_{\sigma}\left\langle
c_{k\sigma}^{\dagger}c_{k\sigma}^{\phantom\dagger}\right\rangle,\\
 n^f(k)=\frac{1}{2}\sum_{\sigma}\left\langle
f_{k\sigma}^{\dagger}f_{k\sigma}^{\phantom\dagger}\right\rangle.
\end{align}
Our DMRG calculation was performed in real space, therefore these quantities can be obtained by
Fourier transforming the corresponding single particle density matrices, namely:
\begin{align}
 n^c(k)=\frac{1}{2}\sum_{jl\sigma}e^{ik(j-l)}\left\langle c_{j\sigma}^{\dagger}c_{l\sigma}^{
\phantom\dagger}\right\rangle,\\
 n^f(k)=\frac{1}{2}\sum_{jl\sigma}e^{ik(j-l)}\left\langle f_{j\sigma}^{\dagger}f_{l\sigma}^{
\phantom\dagger}\right\rangle,
\end{align}
where $k=2\pi n/L $ and $n=-L/2-1,\dots,L/2$. In our case these are symmetric functions, therefore
we consider only the nonnegative $k$ values.
\par
\begin{figure}[!htb]
\includegraphics[width=0.75\columnwidth]{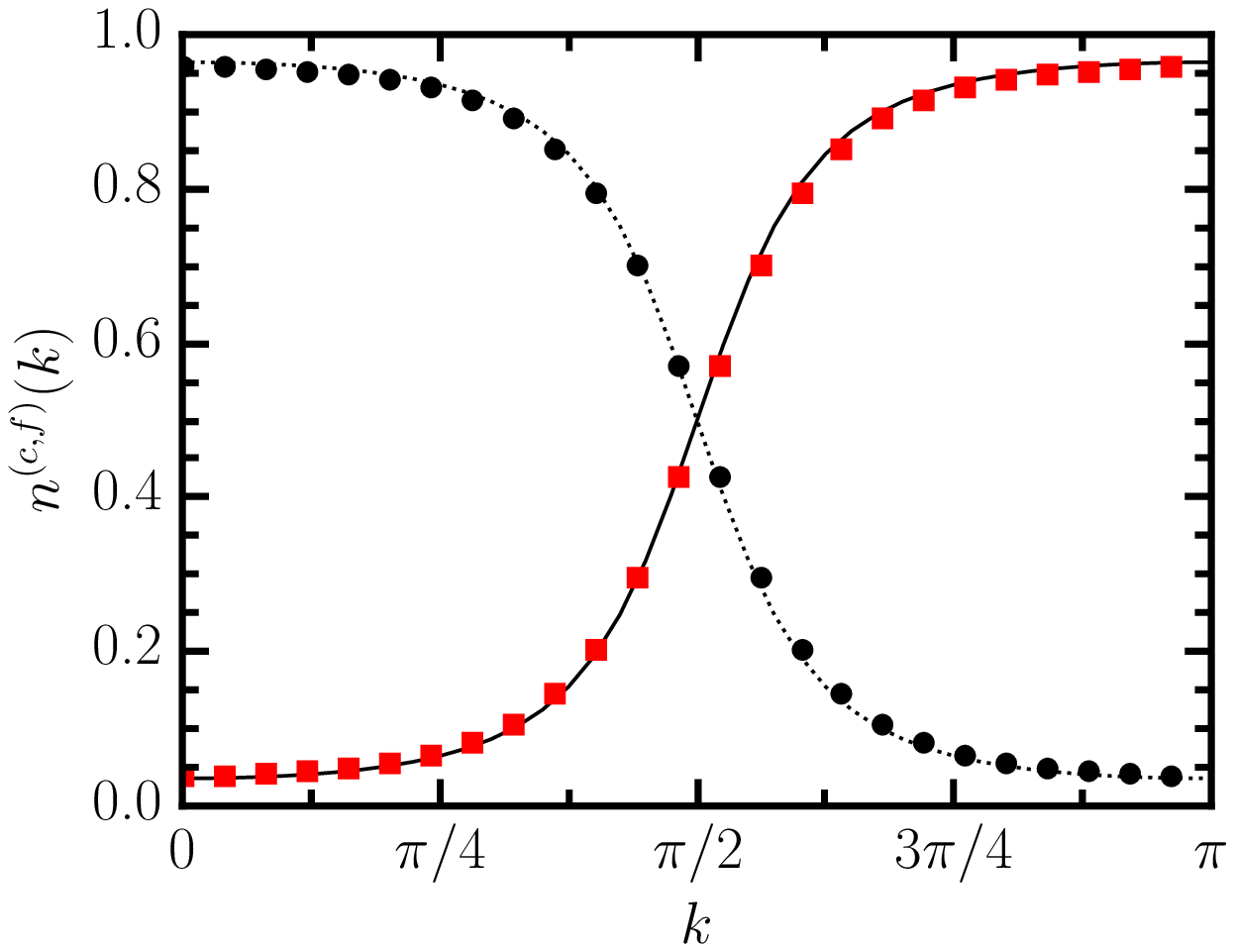}
\caption{Momentum distribution functions of the conduction and $f$
electrons in the noninteracting case ($U_f=U_{cf}=0$,  $V/W=0.1$, $\varepsilon_f=0$). The solid and
dotted lines are obtained from Eq. (\ref{eq:nfk_Uf0}) and (\ref{eq:ndk_Uf0}), respectively. The
symbols \textbullet, \textcolor{red}{$\blacksquare$} denote the DMRG results for $L=50$.
}
\label{fig:mom_dist_Uf0}
\end{figure}

 Before going into the details of the numerical results, we briefly recall the case when
$U_f=U_{cf}=0$ which is easily solvable. Thereby the Hamiltonian can be diagonalized by an unitary
transformation
\begin{equation}
\begin{split}
  \alpha_k^{(-)}&=-v_kc_k+u_kf_k,\\
 \alpha_k^{(+)}&=u_kc_k+v_kf_k,
\end{split}
\end{equation}
where $\alpha_k^{(-)}$ ($\alpha_k^{(+)}$) creates a quasiparticle in the lower (upper) hybridized
band with mixing amplitudes:
\begin{align}\label{eq:nfk_Uf0}
 u_k^2=\frac{1}{2}\left[1-\frac{\varepsilon_k-\varepsilon_f}{\sqrt{
(\varepsilon_k-\varepsilon_f)^2+4V^2}}\right],\\
 v_k^2=\frac{1}{2}\left[1+\frac{\varepsilon_k-\varepsilon_f}{\sqrt{
(\varepsilon_k-\varepsilon_f)^2+4V^2}}\right],
\label{eq:ndk_Uf0}
\end{align}
in our case $\varepsilon_k=-2t\cos k$ and the Fermi momentum is at
the boundary of the Brillouin zone since the lower band is completely filled.
It is easily seen that the momentum distribution functions provide information about the mixing
amplitudes, namely, the portion of conduction and $f$ states in the hybridized band:
\begin{equation}
 \begin{split}
  n^c(k)=v_k^2,\\
  n^f(k)=u_k^2.
 \end{split}
\end{equation}
The momentum
distribution in the noninteracting system is shown in Fig. \ref{fig:mom_dist_Uf0} compared with the
DMRG results. The small discrepancy between the two results is attributed to the open boundary
condition used in DMRG.
In the following we investigate how the interactions modify the above results, using again the DMRG
method with open boundary condition. We checked for short systems that the momentum distributions
of the interacting system calculated with periodic and open boundary conditions are in good
agreement within our error margin.
\par
\begin{figure}[!htb]
\includegraphics[width=\columnwidth]{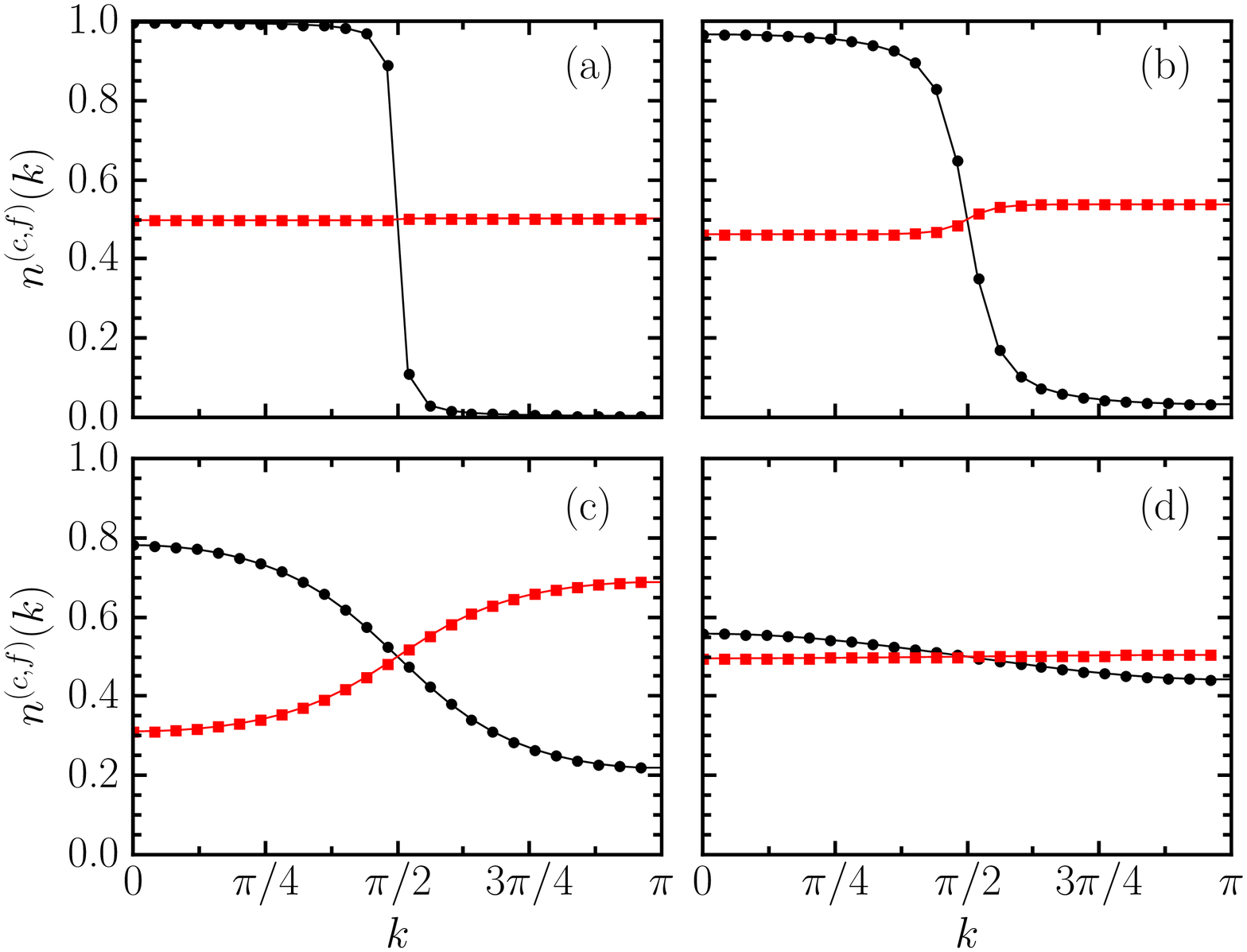}
\caption{Momentum distribution functions of the conduction (\textbullet) and $f$
(\textcolor{red}{$\blacksquare$}) electrons for $L=50$ and $n=2$. Panel (a), (b), (c) and (d)
correspond to $U_{cf}/W=0,1.5,1.7$ and 3, furthermore $U_f/W=3$, $V/W=0.1$ and
$\varepsilon_f=-U_f/2$ in all cases. The lines are guides to the eye.}
\label{fig:mom_dist_half_filled}
\end{figure}

 It has been pointed
out\cite{Shiba:VMC,Hagymasi:Ucf} that strong $U_f$ leads to localization of the $f$ electrons,
since the
number of doubly occupied $f$ levels is negligible and the ground state is a collective singlet.
This is what
we see in Fig. \ref{fig:mom_dist_half_filled} (a), namely, $n^f(k)$ hardly depends on $k$ in the
Kondo regime, while the distribution of
conduction electrons is just slightly affected by $U_f$.
The quantum information analysis\cite{Hagymasi:Ucf} showed that as $U_{cf}$ is switched on more and
more doubly occupied $f$ sites are created,
therefore the $f$ electrons become less localized in real space. Finally, when $U_{cf}$ is large
the $c$ and $f$ electrons tend to avoid each other and the sites are occupied by
two $c$ or two $f$ electrons in an alternating fashion. Now we examine how these features are
reflected in the momentum distributions.
As $U_{cf}$ is switched on, the wave number dependence of the conduction electrons become weaker
and weaker and therefore less
itinerant in real space, while the distribution of the $f$ electrons becomes more and
more dispersive as it can be seen in Fig.
\ref{fig:mom_dist_half_filled} (b) and (c). Above $U_{cf}\approx U_f/2+W/4$ both distributions
hardly depend on the wave number as it is observed in Fig. \ref{fig:mom_dist_half_filled} (d). 
That is, the behavior of the momentum distributions agrees well with the results of the entropy
analysis.

\section{Away from half filling}
\begin{figure}[!htb]
\includegraphics[width=0.75\columnwidth]{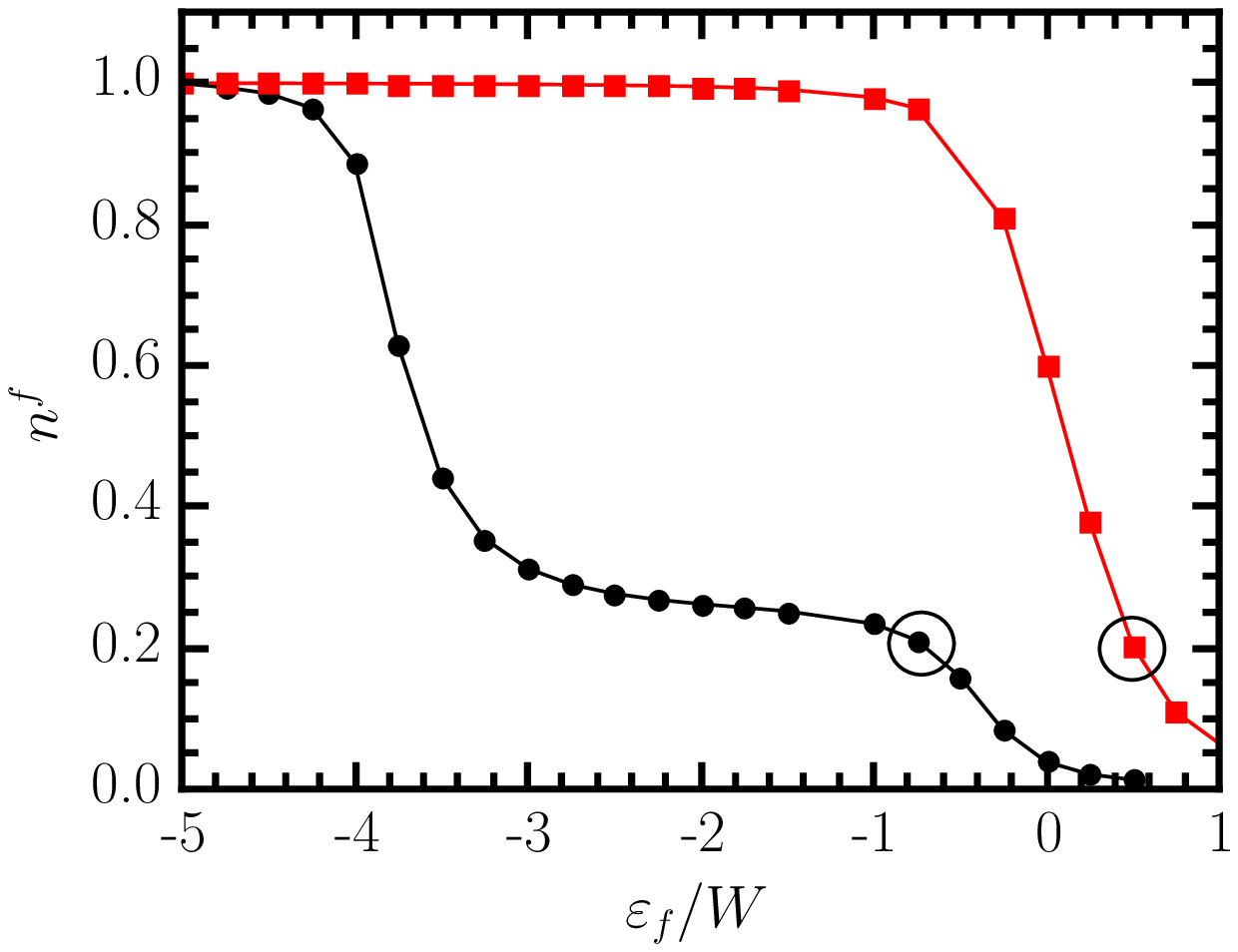}
\caption{The $f$-level occupancy as a function of $\varepsilon_f$ for $L=80$, $U_{cf}=0$
(\textcolor{red}{$\blacksquare$}) and $U_{cf}/W=4$ (\textbullet), furthermore $U_f/W=10$,
$V/W=0.2$. The circled data points are used in the comparison in Fig. \ref{fig:mom_dist_n1p75}. The
lines are guides to the eye.}
\label{fig:nf_ef_n1p75}
\end{figure}

\begin{figure}[!htb]
\includegraphics[width=0.75\columnwidth]{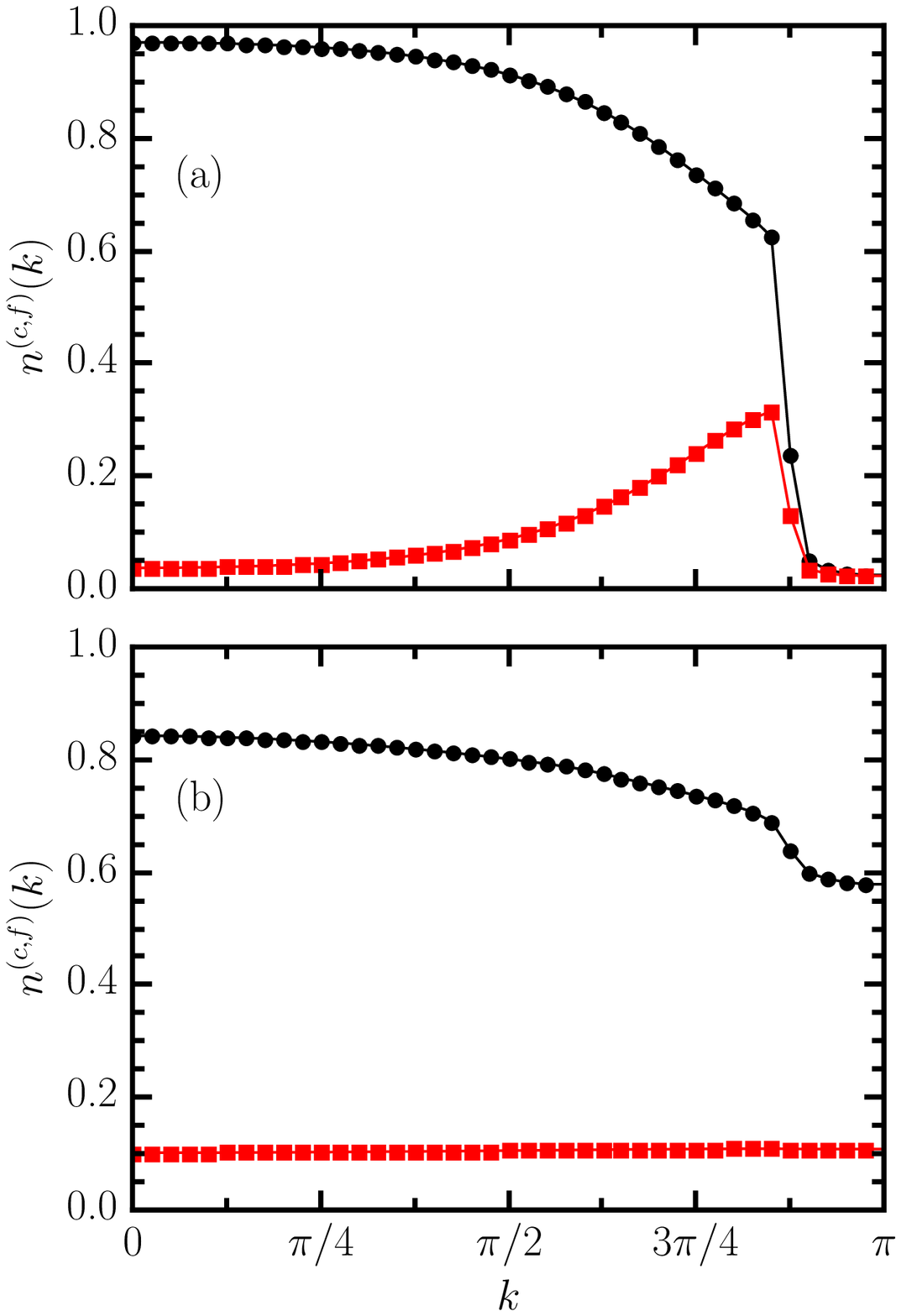}
\caption{Momentum distribution functions of the conduction (\textbullet) and $f$
(\textcolor{red}{$\blacksquare$}) electrons for $L=80$ and $n=1.75$. Panel (a), (b) correspond to
$U_{cf}/W=0$, $\varepsilon_f/W=0.5$ and $U_{cf}/W=4$, $\varepsilon_f/W=-0.75$, respectively,
furthermore $U_f/W=10$, $V/W=0.2$ in all
cases. The lines are guides to the eye. }
\label{fig:mom_dist_n1p75}
\end{figure}

In the previous section we considered the half-filled case. Now we discuss what happens when the
ground state is metallic, and fix the electron density at $n=1.75$. It has been
shown,\cite{Hirashima:cikk,Kubo:GW} that a stable mixed valence regime appears around $n^f=2-n$ in
the presence of $U_{cf}$, and strong enough $U_{cf}$ leads to a first-order transition between the
Kondo and mixed valence states as $\varepsilon_f$ is varied. One can observe in Fig.
\ref{fig:nf_ef_n1p75}, that for strong $U_{cf}$ a stable mixed valence regime appears indeed, and
the change of the valence becomes sharp.
In the following we investigate how the momentum distribution of the electrons change due to
$U_{cf}$ in the mixed valence regime. This is shown in Fig. \ref{fig:mom_dist_n1p75}, where
$\varepsilon_f$,  indicated by the circles around the data points in Fig. \ref{fig:nf_ef_n1p75}, was
chosen such that the occupancy of the $f$ level is nearly the same in the two cases.
One can clearly see that both $n^{c}(k)$ and $n^f(k)$ change drastically around the Fermi
momentum when $U_{cf}=0$. This is not surprising since our system is a Luttinger liquid, where a
logarithmic singularity is expected to occur at the Fermi momentum. For a finite $U_{cf}$ a
significant amount of the conduction and $f$ electrons is scattered above the Fermi momentum and the
$k$-dependence of the electron densities is significantly reduced in both cases, which indicates a
strongly correlated mixed valence state. It is worth noting that the distribution functions do  not
tend to zero above the Fermi momentum for strong $U_{cf}$. These results agree well with what has
been obtained in
infinite dimensions using the Gutzwiller wave function,\cite{Kubo:GW} the main difference is the
absence of the discontinuity at the Fermi momentum due to the one-dimensional property of the model.
\section{Conclusions}
We have investigated an extended periodic Anderson model with an additional Coulomb interaction
using the DMRG algorithm to better understand its effect on the momentum distribution of the
electrons. In
the half-filled, symmetric model (in the Kondo regime), switching on $U_{cf}$ results in the
increased itinerancy of the $f$ electrons, however, above a certain value of $U_{cf}$ it tends to
localize $f$ electrons again. The itinerancy of the conduction electrons is gradually reduced as
$U_{cf}$ is increased. These results agree well with what has been obtained by quantum  information
analysis.\cite{Hagymasi:Ucf} We also investigated what happens when the system is less than
half-filled, that is the
ground state is metallic. It has been revealed that in the mixed valence regime $U_{cf}$ makes both
the conduction and $f$ electrons more correlated. These findings agree qualitatively well with the
properties of the infinite dimensional model, although, as expected for a one-dimensional model,
there is no sharp Fermi edge. This could be analyzed further using the
momentum space version of the DMRG
method.\cite{Xiang:kDMRG,Nishimoto:kDMRG,legeza2003b,georg:unpublished}

\acknowledgments{This work was supported in part by the
Hungarian Research Fund (OTKA) through Grant Nos.~K 100908 and NN110360. }

\bibliography{hund_refs.bib} 

\begin{thebibliography}{10}%
\makeatletter
\providecommand \@ifxundefined [1]{%
 \ifx #1\undefined \expandafter \@firstoftwo
 \else \expandafter \@secondoftwo
\fi
}%
\providecommand \@ifnum [1]{%
 \ifnum #1\expandafter \@firstoftwo
 \else \expandafter \@secondoftwo
\fi
}%
\providecommand \enquote [1]{``#1''}%
\providecommand \bibnamefont  [1]{#1}%
\providecommand \bibfnamefont [1]{#1}%
\providecommand \citenamefont [1]{#1}%
\providecommand\href[0]{\@sanitize\@href}%
\providecommand\@href[1]{\endgroup\@@startlink{#1}\endgroup\@@href}%
\providecommand\@@href[1]{#1\@@endlink}%
\providecommand \@sanitize [0]{\begingroup\catcode`\&12\catcode`\#12\relax}%
\@ifxundefined \pdfoutput {\@firstoftwo}{%
 \@ifnum{\z@=\pdfoutput}{\@firstoftwo}{\@secondoftwo}%
}{%
 \providecommand\@@startlink[1]{\leavevmode\special{html:<a href="#1">}}%
 \providecommand\@@endlink[0]{\special{html:</a>}}%
}{%
 \providecommand\@@startlink[1]{%
  \leavevmode
  \pdfstartlink
   attr{/Border[0 0 1 ]/H/I/C[0 1 1]}%
   user{/Subtype/Link/A<</Type/Action/S/URI/URI(#1)>>}%
  \relax
 }%
 \providecommand\@@endlink[0]{\pdfendlink}%
}%
\providecommand \url  [0]{\begingroup\@sanitize \@url }%
\providecommand \@url [1]{\endgroup\@href {#1}{\urlprefix}}%
\providecommand \urlprefix [0]{URL }%
\providecommand \Eprint[0]{\href }%
\@ifxundefined \urlstyle {%
  \providecommand \doi [1]{doi:\discretionary{}{}{}#1}%
}{%
  \providecommand \doi [0]{doi:\discretionary{}{}{}\begingroup
  \urlstyle{rm}\Url }%
}%
\providecommand \doibase [0]{http://dx.doi.org/}%
\providecommand \Doi[1]{\href{\doibase#1}}%
\providecommand \bibAnnote [3]{%
  \BibitemShut{#1}%
  \begin{quotation}\noindent
    \textsc{Key:}\ #2\\\textsc{Annotation:}\ #3%
  \end{quotation}%
}%
\providecommand \bibAnnoteFile [2]{%
  \IfFileExists{#2}{\bibAnnote {#1} {#2} {\input{#2}}}{}%
}%
\providecommand \typeout [0]{\immediate \write \m@ne }%
\providecommand \selectlanguage [0]{\@gobble}%
\providecommand \bibinfo [0]{\@secondoftwo}%
\providecommand \bibfield [0]{\@secondoftwo}%
\providecommand \translation [1]{[#1]}%
\providecommand \BibitemOpen[0]{}%
\providecommand \bibitemStop [0]{}%
\providecommand \bibitemNoStop [0]{.\EOS\space}%
\providecommand \EOS [0]{\spacefactor3000\relax}%
\providecommand \BibitemShut [1]{\csname bibitem#1\endcsname}%
\bibitem{Hewson:book}%
  \BibitemOpen
  \bibfield{author}{%
  \bibinfo {author} {\bibfnamefont{A.~C.}\ \bibnamefont{Hewson}},\ }%
  \emph{\bibinfo {title} {The Kondo Problem to Heavy Fermions}}\ (\bibinfo
  {publisher} {Cambridge University Press},\ \bibinfo {address} {Cambridge},\
  \bibinfo {year} {1993})%
  \bibAnnoteFile{NoStop}{Hewson:book}%
\bibitem{Patrik:book}%
  \BibitemOpen
  \bibfield{author}{%
  \bibinfo {author} {\bibfnamefont{P.}~\bibnamefont{Fazekas}},\ }%
  \emph{\bibinfo {title} {Lecture notes on electron correlation and
  magnetism}}\ (\bibinfo {publisher} {World Scientific},\ \bibinfo {address}
  {Singapore},\ \bibinfo {year} {1999})%
  \bibAnnoteFile{NoStop}{Patrik:book}%
\bibitem{Yuan:science}%
  \BibitemOpen
  \bibfield{author}{%
  \bibinfo {author} {\bibfnamefont{H.~Q.}\ \bibnamefont{Yuan}}, \bibinfo
  {author} {\bibfnamefont{F.~M.}\ \bibnamefont{Grosche}}, \bibinfo {author}
  {\bibfnamefont{M.}~\bibnamefont{Deppe}}, \bibinfo {author}
  {\bibfnamefont{C.}~\bibnamefont{Geibel}}, \bibinfo {author}
  {\bibfnamefont{G.}~\bibnamefont{Sparn}},\ and\ \bibinfo {author}
  {\bibfnamefont{F.}~\bibnamefont{Steglich}},\ }%
  \bibfield{journal}{%
  \bibinfo {journal} {Science}\ }%
  \textbf{\bibinfo {volume} {302}},\ \bibinfo {pages} {2104} (\bibinfo {year}
  {2003})%
  \bibAnnoteFile{NoStop}{Yuan:science}%
\bibitem{CeCu2Si2:meres_2}%
  \BibitemOpen
  \bibfield{author}{%
  \bibinfo {author} {\bibfnamefont{A.~T.}\ \bibnamefont{Holmes}}, \bibinfo
  {author} {\bibfnamefont{D.}~\bibnamefont{Jaccard}},\ and\ \bibinfo {author}
  {\bibfnamefont{K.}~\bibnamefont{Miyake}},\ }%
  \bibfield{journal}{%
  \bibinfo {journal} {Phys. Rev. B}\ }%
  \textbf{\bibinfo {volume} {69}},\ \bibinfo {pages} {024508} (\bibinfo {year}
  {2004})%
  \bibAnnoteFile{NoStop}{CeCu2Si2:meres_2}%
\bibitem{CeCu2SiGe:meres}%
  \BibitemOpen
  \bibfield{author}{%
  \bibinfo {author} {\bibfnamefont{H.~Q.}\ \bibnamefont{Yuan}}, \bibinfo
  {author} {\bibfnamefont{F.~M.}\ \bibnamefont{Grosche}}, \bibinfo {author}
  {\bibfnamefont{M.}~\bibnamefont{Deppe}}, \bibinfo {author}
  {\bibfnamefont{G.}~\bibnamefont{Sparn}}, \bibinfo {author}
  {\bibfnamefont{C.}~\bibnamefont{Geibel}},\ and\ \bibinfo {author}
  {\bibfnamefont{F.}~\bibnamefont{Steglich}},\ }%
  \bibfield{journal}{%
  \bibinfo {journal} {Phys. Rev. Lett.}\ }%
  \textbf{\bibinfo {volume} {96}},\ \bibinfo {pages} {047008} (\bibinfo {year}
  {2006})%
  \bibAnnoteFile{NoStop}{CeCu2SiGe:meres}%
\bibitem{CeCu2Si2:meres_1}%
  \BibitemOpen
  \bibfield{author}{%
  \bibinfo {author} {\bibfnamefont{K.}~\bibnamefont{Fujiwara}}, \bibinfo
  {author} {\bibfnamefont{Y.}~\bibnamefont{Hata}}, \bibinfo {author}
  {\bibfnamefont{K.}~\bibnamefont{Kobayashi}}, \bibinfo {author}
  {\bibfnamefont{K.}~\bibnamefont{Miyoshi}}, \bibinfo {author}
  {\bibfnamefont{J.}~\bibnamefont{Takeuchi}}, \bibinfo {author}
  {\bibfnamefont{Y.}~\bibnamefont{Shimaoka}}, \bibinfo {author}
  {\bibfnamefont{H.}~\bibnamefont{Kotegawa}}, \bibinfo {author}
  {\bibfnamefont{T.~C.}\ \bibnamefont{Kobayashi}}, \bibinfo {author}
  {\bibfnamefont{C.}~\bibnamefont{Geibel}},\ and\ \bibinfo {author}
  {\bibfnamefont{F.}~\bibnamefont{Steglich}},\ }%
  \bibfield{journal}{%
  \bibinfo {journal} {J. Phys. Soc. Jpn.}\ }%
  \textbf{\bibinfo {volume} {77}},\ \bibinfo {pages} {123711} (\bibinfo {year}
  {2008})%
  \bibAnnoteFile{NoStop}{CeCu2Si2:meres_1}%
\bibitem{CeCu2Si2:meres_SC}%
  \BibitemOpen
  \bibfield{author}{%
  \bibinfo {author} {\bibfnamefont{E.}~\bibnamefont{Lengyel}}, \bibinfo
  {author} {\bibfnamefont{M.}~\bibnamefont{Nicklas}}, \bibinfo {author}
  {\bibfnamefont{H.~S.}\ \bibnamefont{Jeevan}}, \bibinfo {author}
  {\bibfnamefont{G.}~\bibnamefont{Sparn}}, \bibinfo {author}
  {\bibfnamefont{C.}~\bibnamefont{Geibel}}, \bibinfo {author}
  {\bibfnamefont{F.}~\bibnamefont{Steglich}}, \bibinfo {author}
  {\bibfnamefont{Y.}~\bibnamefont{Yoshioka}},\ and\ \bibinfo {author}
  {\bibfnamefont{K.}~\bibnamefont{Miyake}},\ }%
  \bibfield{journal}{%
  \bibinfo {journal} {Phys. Rev. B}\ }%
  \textbf{\bibinfo {volume} {80}},\ \bibinfo {pages} {140513} (\bibinfo {year}
  {2009})%
  \bibAnnoteFile{NoStop}{CeCu2Si2:meres_SC}%
\bibitem{RueffPRL_meres:cikk}%
  \BibitemOpen
  \bibfield{author}{%
  \bibinfo {author} {\bibfnamefont{J.-P.}\ \bibnamefont{Rueff}}, \bibinfo
  {author} {\bibfnamefont{S.}~\bibnamefont{Raymond}}, \bibinfo {author}
  {\bibfnamefont{M.}~\bibnamefont{Taguchi}}, \bibinfo {author}
  {\bibfnamefont{M.}~\bibnamefont{Sikora}}, \bibinfo {author}
  {\bibfnamefont{J.-P.}\ \bibnamefont{Iti\'e}}, \bibinfo {author}
  {\bibfnamefont{F.}~\bibnamefont{Baudelet}}, \bibinfo {author}
  {\bibfnamefont{D.}~\bibnamefont{Braithwaite}}, \bibinfo {author}
  {\bibfnamefont{G.}~\bibnamefont{Knebel}},\ and\ \bibinfo {author}
  {\bibfnamefont{D.}~\bibnamefont{Jaccard}},\ }%
  \bibfield{journal}{%
  \bibinfo {journal} {Phys. Rev. Lett.}\ }%
  \textbf{\bibinfo {volume} {106}},\ \bibinfo {pages} {186405} (\bibinfo {year}
  {2011})%
  \bibAnnoteFile{NoStop}{RueffPRL_meres:cikk}%
\bibitem{Miyake:VMC}%
  \BibitemOpen
  \bibfield{author}{%
  \bibinfo {author} {\bibfnamefont{Y.}~\bibnamefont{Onishi}}\ and\ \bibinfo
  {author} {\bibfnamefont{K.}~\bibnamefont{Miyake}},\ }%
  \bibfield{journal}{%
  \bibinfo {journal} {J. Phys. Soc. Jpn.}\ }%
  \textbf{\bibinfo {volume} {69}},\ \bibinfo {pages} {3955} (\bibinfo {year}
  {2000})%
  \bibAnnoteFile{NoStop}{Miyake:VMC}%
\bibitem{Miyake:CVF_1}%
  \BibitemOpen
  \bibfield{author}{%
  \bibinfo {author} {\bibfnamefont{K.}~\bibnamefont{Miyake}}\ and\ \bibinfo
  {author} {\bibfnamefont{H.}~\bibnamefont{Maebashi}},\ }%
  \bibfield{journal}{%
  \bibinfo {journal} {J. Phys. Soc. Jpn.}\ }%
  \textbf{\bibinfo {volume} {71}},\ \bibinfo {pages} {1007} (\bibinfo {year}
  {2002})%
  \bibAnnoteFile{NoStop}{Miyake:CVF_1}%
\bibitem{DMRG:Miyake1}%
  \BibitemOpen
  \bibfield{author}{%
  \bibinfo {author} {\bibfnamefont{S.}~\bibnamefont{Watanabe}}, \bibinfo
  {author} {\bibfnamefont{M.}~\bibnamefont{Imada}},\ and\ \bibinfo {author}
  {\bibfnamefont{K.}~\bibnamefont{Miyake}},\ }%
  \bibfield{journal}{%
  \bibinfo {journal} {J. Phys. Soc. Jpn.}\ }%
  \textbf{\bibinfo {volume} {75}},\ \bibinfo {pages} {043710} (\bibinfo {year}
  {2006})%
  \bibAnnoteFile{NoStop}{DMRG:Miyake1}%
\bibitem{DMRG:Miyake2}%
  \BibitemOpen
  \bibfield{author}{%
  \bibinfo {author} {\bibfnamefont{S.}~\bibnamefont{Watanabe}}, \bibinfo
  {author} {\bibfnamefont{M.}~\bibnamefont{Imada}},\ and\ \bibinfo {author}
  {\bibfnamefont{K.}~\bibnamefont{Miyake}},\ }%
  \bibfield{journal}{%
  \bibinfo {journal} {J. Magn. and Magn. Mat.}\ }%
  \textbf{\bibinfo {volume} {310}} (\bibinfo {year} {2007})%
  \bibAnnoteFile{NoStop}{DMRG:Miyake2}%
\bibitem{Miyake_SC}%
  \BibitemOpen
  \bibfield{author}{%
  \bibinfo {author} {\bibfnamefont{A.~T.}\ \bibnamefont{Holmes}}, \bibinfo
  {author} {\bibfnamefont{D.}~\bibnamefont{Jaccard}},\ and\ \bibinfo {author}
  {\bibfnamefont{K.}~\bibnamefont{Miyake}},\ }%
  \bibfield{journal}{%
  \bibinfo {journal} {J. Phys. Soc. Jpn.}\ }%
  \textbf{\bibinfo {volume} {76}},\ \bibinfo {pages} {051002} (\bibinfo {year}
  {2007})%
  \bibAnnoteFile{NoStop}{Miyake_SC}%
\bibitem{Miyake:review}%
  \BibitemOpen
  \bibfield{author}{%
  \bibinfo {author} {\bibfnamefont{K.}~\bibnamefont{Miyake}},\ }%
  \bibfield{journal}{%
  \bibinfo {journal} {J. Phys.: Condens. Matter}\ }%
  \textbf{\bibinfo {volume} {19}},\ \bibinfo {pages} {125201} (\bibinfo {year}
  {2007})%
  \bibAnnoteFile{NoStop}{Miyake:review}%
\bibitem{Hirashima:cikk}%
  \BibitemOpen
  \bibfield{author}{%
  \bibinfo {author} {\bibfnamefont{Y.}~\bibnamefont{Saiga}}, \bibinfo {author}
  {\bibfnamefont{T.}~\bibnamefont{Sugibayashi}},\ and\ \bibinfo {author}
  {\bibfnamefont{D.~S.}\ \bibnamefont{Hirashima}},\ }%
  \bibfield{journal}{%
  \bibinfo {journal} {J. Phys. Soc. Jpn.}\ }%
  \textbf{\bibinfo {volume} {77}},\ \bibinfo {pages} {114710} (\bibinfo {year}
  {2008})%
  \bibAnnoteFile{NoStop}{Hirashima:cikk}%
\bibitem{Kubo:GW}%
  \BibitemOpen
  \bibfield{author}{%
  \bibinfo {author} {\bibfnamefont{K.}~\bibnamefont{Kubo}},\ }%
  \bibfield{journal}{%
  \bibinfo {journal} {J. Phys. Soc. Jpn.}\ }%
  \textbf{\bibinfo {volume} {80}},\ \bibinfo {pages} {114711} (\bibinfo {year}
  {2011})%
  \bibAnnoteFile{NoStop}{Kubo:GW}%
\bibitem{Hagymasi:GW}%
  \BibitemOpen
  \bibfield{author}{%
  \bibinfo {author} {\bibfnamefont{I.}~\bibnamefont{Hagym\'asi}}, \bibinfo
  {author} {\bibfnamefont{K.}~\bibnamefont{Itai}},\ and\ \bibinfo {author}
  {\bibfnamefont{J.}~\bibnamefont{S\'olyom}},\ }%
  \bibfield{journal}{%
  \bibinfo {journal} {Phys. Rev. B}\ }%
  \textbf{\bibinfo {volume} {87}},\ \bibinfo {pages} {125146} (\bibinfo {year}
  {2013})%
  \bibAnnoteFile{NoStop}{Hagymasi:GW}%
\bibitem{Kawakami:DMFT_1}%
  \BibitemOpen
  \bibfield{author}{%
  \bibinfo {author} {\bibfnamefont{T.}~\bibnamefont{Yoshida}}, \bibinfo
  {author} {\bibfnamefont{T.}~\bibnamefont{Ohashi}},\ and\ \bibinfo {author}
  {\bibfnamefont{N.}~\bibnamefont{Kawakami}},\ }%
  \bibfield{journal}{%
  \bibinfo {journal} {J. Phys. Soc. Jpn.}\ }%
  \textbf{\bibinfo {volume} {80}},\ \bibinfo {pages} {064710} (\bibinfo {year}
  {2011})%
  \bibAnnoteFile{NoStop}{Kawakami:DMFT_1}%
\bibitem{Kawakami:DMFT_2}%
  \BibitemOpen
  \bibfield{author}{%
  \bibinfo {author} {\bibfnamefont{T.}~\bibnamefont{Yoshida}}\ and\ \bibinfo
  {author} {\bibfnamefont{N.}~\bibnamefont{Kawakami}},\ }%
  \bibfield{journal}{%
  \bibinfo {journal} {Phys. Rev. B}\ }%
  \textbf{\bibinfo {volume} {85}},\ \bibinfo {pages} {235148} (\bibinfo {year}
  {2012})%
  \bibAnnoteFile{NoStop}{Kawakami:DMFT_2}%
\bibitem{Hagymasi:Ucf}%
  \BibitemOpen
  \bibfield{author}{%
  \bibinfo {author} {\bibfnamefont{I.}~\bibnamefont{Hagym\'asi}}, \bibinfo
  {author} {\bibfnamefont{J.}~\bibnamefont{S\'olyom}},\ and\ \bibinfo {author}
  {\bibfnamefont{{\"O}.}~\bibnamefont{Legeza}},\ }%
  \bibfield{journal}{%
  \bibinfo {journal} {Phys. Rev. B}\ }%
  \textbf{\bibinfo {volume} {90}},\ \bibinfo {pages} {125137} (\bibinfo {year}
  {2014})%
  \bibAnnoteFile{NoStop}{Hagymasi:Ucf}%
\bibitem{White:DMRG1}%
  \BibitemOpen
  \bibfield{author}{%
  \bibinfo {author} {\bibfnamefont{S.~R.}\ \bibnamefont{White}},\ }%
  \bibfield{journal}{%
  \bibinfo {journal} {Phys. Rev. Lett.}\ }%
  \textbf{\bibinfo {volume} {69}},\ \bibinfo {pages} {2863} (\bibinfo {year}
  {1992})%
  \bibAnnoteFile{NoStop}{White:DMRG1}%
\bibitem{White:DMRG2}%
  \BibitemOpen
  \bibfield{author}{%
  \bibinfo {author} {\bibfnamefont{S.~R.}\ \bibnamefont{White}},\ }%
  \bibfield{journal}{%
  \bibinfo {journal} {Phys. Rev. B}\ }%
  \textbf{\bibinfo {volume} {48}},\ \bibinfo {pages} {10345} (\bibinfo {year}
  {1993})%
  \bibAnnoteFile{NoStop}{White:DMRG2}%
\bibitem{schollwock2005}%
  \BibitemOpen
  \bibfield{author}{%
  \bibinfo {author} {\bibfnamefont{U.}~\bibnamefont{Schollw\"ock}},\ }%
  \bibfield{journal}{%
  \bibinfo {journal} {Rev. Mod. Phys.}\ }%
  \textbf{\bibinfo {volume} {77}},\ \bibinfo {pages} {259} (\bibinfo {year}
  {2005})%
  \bibAnnoteFile{NoStop}{schollwock2005}%
\bibitem{manmana2005}%
  \BibitemOpen
  \bibfield{author}{%
  \bibinfo {author} {\bibfnamefont{R.~M.}\ \bibnamefont{Noack}}\ and\ \bibinfo
  {author} {\bibfnamefont{S.~R.}\ \bibnamefont{Manmana}},\ }%
  \bibfield{journal}{%
  \bibinfo {journal} {AIP Conf. Proc.}\ }%
  \textbf{\bibinfo {volume} {789}},\ \bibinfo {pages} {93} (\bibinfo {year}
  {2005})%
  \bibAnnoteFile{NoStop}{manmana2005}%
\bibitem{hallberg2006}%
  \BibitemOpen
  \bibfield{author}{%
  \bibinfo {author} {\bibfnamefont{K.}~\bibnamefont{Hallberg}},\ }%
  \bibfield{journal}{%
  \bibinfo {journal} {Adv. Phys.}\ }%
  \textbf{\bibinfo {volume} {55}},\ \bibinfo {pages} {477} (\bibinfo {year}
  {2006})%
  \bibAnnoteFile{NoStop}{hallberg2006}%
\bibitem{DBSS:cikk1}%
  \BibitemOpen
  \bibfield{author}{%
  \bibinfo {author} {\bibfnamefont{{\"O}.}~\bibnamefont{Legeza}}, \bibinfo
  {author} {\bibfnamefont{J.}~\bibnamefont{R\"oder}},\ and\ \bibinfo {author}
  {\bibfnamefont{B.~A.}\ \bibnamefont{Hess}},\ }%
  \bibfield{journal}{%
  \bibinfo {journal} {Phys. Rev. B}\ }%
  \textbf{\bibinfo {volume} {67}},\ \bibinfo {pages} {125114} (\bibinfo {year}
  {2003})%
  \bibAnnoteFile{NoStop}{DBSS:cikk1}%
\bibitem{DBSS:cikk2}%
  \BibitemOpen
  \bibfield{author}{%
  \bibinfo {author} {\bibfnamefont{{\"O}.}~\bibnamefont{Legeza}}\ and\ \bibinfo
  {author} {\bibfnamefont{J.}~\bibnamefont{S\'olyom}},\ }%
  \bibfield{journal}{%
  \bibinfo {journal} {Phys. Rev. B}\ }%
  \textbf{\bibinfo {volume} {70}},\ \bibinfo {pages} {205118} (\bibinfo {year}
  {2004})%
  \bibAnnoteFile{NoStop}{DBSS:cikk2}%
\bibitem{Shiba:VMC}%
  \BibitemOpen
  \bibfield{author}{%
  \bibinfo {author} {\bibfnamefont{H.}~\bibnamefont{Shiba}},\ }%
  \bibfield{journal}{%
  \bibinfo {journal} {J. Phys. Soc. Jpn.}\ }%
  \textbf{\bibinfo {volume} {55}},\ \bibinfo {pages} {2765} (\bibinfo {year}
  {1986})%
  \bibAnnoteFile{NoStop}{Shiba:VMC}%
\bibitem{Xiang:kDMRG}%
  \BibitemOpen
  \bibfield{author}{%
  \bibinfo {author} {\bibfnamefont{T.}~\bibnamefont{Xiang}},\ }%
  \bibfield{journal}{%
  \bibinfo {journal} {Phys. Rev. B}\ }%
  \textbf{\bibinfo {volume} {53}},\ \bibinfo {pages} {R10445} (\bibinfo {year}
  {1996})%
  \bibAnnoteFile{NoStop}{Xiang:kDMRG}%
\bibitem{Nishimoto:kDMRG}%
  \BibitemOpen
  \bibfield{author}{%
  \bibinfo {author} {\bibfnamefont{S.}~\bibnamefont{Nishimoto}}, \bibinfo
  {author} {\bibfnamefont{E.}~\bibnamefont{Jeckelmann}}, \bibinfo {author}
  {\bibfnamefont{F.}~\bibnamefont{Gebhard}},\ and\ \bibinfo {author}
  {\bibfnamefont{R.~M.}\ \bibnamefont{Noack}},\ }%
  \bibfield{journal}{%
  \bibinfo {journal} {Phys. Rev. B}\ }%
  \textbf{\bibinfo {volume} {65}},\ \bibinfo {pages} {165114} (\bibinfo {year}
  {2002})%
  \bibAnnoteFile{NoStop}{Nishimoto:kDMRG}%
\bibitem{legeza2003b}%
  \BibitemOpen
  \bibfield{author}{%
  \bibinfo {author} {\bibfnamefont{{\"O}.}~\bibnamefont{Legeza}}\ and\ \bibinfo
  {author} {\bibfnamefont{J.}~\bibnamefont{S\'olyom}},\ }%
  \bibfield{journal}{%
  \bibinfo {journal} {Phys. Rev. B}\ }%
  \textbf{\bibinfo {volume} {68}},\ \bibinfo {pages} {195116} (\bibinfo {year}
  {2003})%
  \bibAnnoteFile{NoStop}{legeza2003b}%
\bibitem{georg:unpublished}%
  \BibitemOpen
  \bibfield{author}{%
  \bibinfo {author} {\bibfnamefont{G.}~\bibnamefont{Ehlers}}, \bibinfo {author}
  {\bibfnamefont{J.}~\bibnamefont{S\'olyom}}, \bibinfo {author}
  {\bibfnamefont{{\"O}.}~\bibnamefont{Legeza}},\ and\ \bibinfo {author}
  {\bibfnamefont{R.~M.}\ \bibnamefont{Noack}},\ }%
  \bibinfo {journal} {(unpublished)}%
  \bibAnnoteFile{NoStop}{georg:unpublished}%
\end{thebibliography}%

\end{document}